\documentclass[alpha-refs]{wiley-article}

\usepackage{siunitx}
\usepackage{hyperref}
\usepackage{gensymb}

\usepackage{tabularx}
\usepackage{adjustbox}

\usepackage{natbib}
\bibpunct{(}{)}{;}{z}{}{,}

\papertype{Original Article}

\title{Assessing the predictability of Medicanes in ECMWF ensemble forecasts using an object-based approach}

\author[1,2]{Enrico Di Muzio}
\author[2]{Michael Riemer}
\author[1]{Andreas H. Fink}
\author[1]{Michael Maier-Gerber}

\affil[1]{Institute of Meteorology and Climate Research, Karlsruhe Institute of Technology, Karlsruhe, 76131, Germany}
\affil[2]{Institute of Atmospheric Physics, Johannes Gutenberg University Mainz, Mainz, 55128, Germany}

\corraddress{Enrico Di Muzio, Institute of Meteorology and Climate Research, Karlsruhe Institute of Technology, Karlsruhe, 76131, Germany}
\corremail{enrico.dimuzio@kit.edu}

\runningauthor{Di Muzio et al.}

\begin{document}

\maketitle

\begin{abstract}
The predictability of eight southern European tropical-like cyclones, seven of which Medicanes, is studied evaluating ECMWF operational ensemble forecasts against operational analysis data. Forecast cyclone trajectories are compared to the cyclone trajectory in the analysis by means of a dynamic time warping technique, which allows to find a match in terms of their overall spatio-temporal similarity. Each storm is treated as an object and its forecasts are analysed using metrics that describe intensity, symmetry, compactness, and upper-level thermal structure. This object-based approach allows to focus on specific storm features, while tolerating their shifts in time and space to some extent.

The compactness and symmetry of the storms are generally underpredicted, especially at long lead times. However, forecast accuracy tends to strongly improve at short lead times, indicating that the ECMWF ensemble forecast model can adequately reproduce Medicanes, albeit only few days in advance. In particular, late forecasts which have been initialised when the cyclone has already developed are distinctly more accurate than earlier forecasts in predicting its kinematic and thermal structure, confirming previous findings of high sensitivity of Medicane simulations to initial conditions.

Findings reveal a markedly non-gradual evolution of ensemble forecasts with lead time, which is often far from a progressive convergence towards the analysis value. Specifically, a rapid increase in the probability of cyclone occurrence (``forecast jump'') is seen in most cases, generally between 5 and 7 days lead time. Jumps are also found for ensemble median and/or spread for storm thermal structure forecasts. This behaviour is compatible with the existence of predictability barriers. On the other hand, storm position forecasts often exhibit a consistent spatial distribution of storm position uncertainty and bias between consecutive forecasts.

\end{abstract}

\section{Introduction} \label{sec:intro}

The Mediterranean region has long been known as a hotspot for cyclogenesis \citep{peterssen_1956} due to its geography \citep{buzzi_1978}. Despite its relatively high latitude, a small but significant fraction of Mediterranean cyclones displays some similarity to tropical cyclones both in their appearance in satellite images and in their kinematic and thermal structure. Such Mediterranean tropical-like cyclones, also known as \emph{Medicanes} (short for Mediterranean hurricanes), have been documented since the beginning of the satellite era \citep{ernst_1983,mayengon_1984,rasmussen_1987}. Medicanes constitute a major threat due to intense winds, torrential rainfall and associated floods. These storms are usually shorter-lived than North Atlantic hurricanes but may exhibit several tropical-like traits in the mature phase of their life cycle, such as high axial symmetry, a warm core, a strong tendency to weaken after making landfall and a cloud-free, weak-wind region at their centre resembling the eye of a hurricane \citep{emanuel_2005,cavicchia_2014_clim}.

Medicanes are distinguished among Mediterranean cyclones by the complex pathway leading to their formation and maintenance. While hurricanes develop in regions of near-zero baroclinicity and draw their energy from very warm tropical oceans, Medicanes arise from pressure lows that are born under moderate to strong baroclinicity. The interaction between the warm sea and cold air associated to a deep upper-level trough provides the necessary thermodynamic disequilibrium for these storms to develop a warm core \citep{emanuel_2005,cavicchia_2014_clim}. It can thus be maintained that Medicanes are the result of a synergy between synoptic-scale processes, which provide the necessary environment for their development, and mesoscale processes such as deep convection and latent heat fluxes from the sea, which are crucial for their maintenance \citep{homar_2003,emanuel_2005,tous_fluxes}. For this reason and because of their small size \citep{miglietta_2013,picornell_2014}, scarce data availability and the complex orography of the Mediterranean region, predictability of Medicanes is generally limited \citep{claud_2010,pantillon_2013}.

Despite the increased research interest in the last two decades, Medicanes are by nature elusive, due to their low frequency of occurrence \citep[less than two per year, according to][]{cavicchia_2014_clim} and the fact that they normally occur over the sea, where observations are sparse. For this reason, many studies so far have focused on modeling aspects \citep{homar_2003,fita_2007,davolio_2009,miglietta_2011,chaboureau_verif_2012,miglietta_2013,cioni_2016,mazza_2017,pytharoulis_al_2017,cioni_2018} and fewer on observational aspects \citep{pytharoulis_2000,reale_2001,moscatello_2008,chaboureau_pantillon_2012,miglietta_2013}. Further studies examined Medicanes in relation to climate change \citep{romero_2013,cavicchia_2014_fut,walsh_2014,romero_2017}, a critical aspect given the vulnerability of the Mediterranean region to future climate change \citep{giorgi_2008}. Research efforts so far focused on deterministic simulations of Medicanes using high-resolution, convection-permitting models \citep{fita_2007,davolio_2009,cioni_2016,mazza_2017,pytharoulis_al_2017,cioni_2018} as they are deemed to best reproduce small-scale processes playing a crucial role in storm maintenance during the tropical-like phase. Few studies analysed Medicanes using ensemble forecasts \citep{cavicchia_2012,chaboureau_verif_2012,pantillon_2013,mazza_2017} of which only \cite{pantillon_2013} used operational ensemble forecasts.

Ensemble forecasts have been shown to be a valuable tool to predict extreme weather events several days in advance \citep[e.g.][]{buizza_2002,palmer_2002,lalaurette_2003,magnusson_2015}, analyse tropical cyclones \citep{torn_2013,rios_2016} and their predictability \citep{munsell_2013,zhang_2013}. Among operational ensemble forecast systems, the European Centre for Medium-Range Weather Forecasts (ECMWF) model has shown high predictive skills for extreme weather events \citep{lalaurette_2003} including tropical cyclones \citep{yamaguchi_2010,hamill_2011} and it has been successfully used to study their predictability \citep{magnusson_2014,gonzalez_2018}. Specifically, \cite{pantillon_2013} used ECMWF operational ensemble forecasts to study the predictability of the 2006 Medicane and found that they were able to more consistently capture early signals of its occurrence with respect to ECMWF deterministic forecasts.

The present study fills a gap in the existing literature in that it systematically investigates the predictability of eight recent (2011-2017) southern European tropical-like cyclones (seven of which are Medicanes), evaluating ECMWF operational ensemble forecasts against operational analysis. Our goal is to assess whether and how long in advance these forecasts can adequately reproduce Medicanes. We also analyse the temporal evolution of the predictability of these storms by identifying particularly rapid changes of the ensemble statistics with lead time (hereafter referred to as \emph{forecast jumps} for brevity) that stand out compared to the gradual convergence towards the analysis value that might be expected. We finally investigate whether there is any consistent bias in the ensemble forecasts, as it could be expected given the model's relatively low horizontal resolution and parameterized convection. This is not a straightforward task, as Medicanes are by their very nature low-probability events and as such are found near the tail of the forecast distribution, as observed by \cite{majumdar_2014}.

In this study, we evaluate ensemble forecasts against analysis data using an object-based approach. Object-based methods gained popularity in recent decades for the verification of precipitation forecasts \citep{ebert_2000,wernli_2008} and have since been applied to the analysis of other atmospheric features, such as the jet stream \citep{limbach_2012} and Rossby waves \citep{wiegand_2014}. These methods allow to avoid the ``double penalty problem'' \citep{ebert_2000,wernli_2008} that arises in case of a mere displacement of an otherwise well predicted atmospheric feature if an Eulerian error metric is used. An object-based method, in contrast, can identify that the ensemble spread is due to a displacement of the feature and that only this aspect exhibits reduced predictability. In addition to spatial displacements, we allow (small) temporal shifts of the forecast features. This approach helps identify matching features in the forecast, which is beneficial to study low-probability weather events, in particular at longer lead times.

The paper is structured as follows. In Section \ref{sec:methods}, the data and methods used are described: in particular, a dynamic time warping technique that allows to consider temporal shifts in the forecasts is illustrated in detail. A brief overview of the eight storms analysed is given in Section \ref{sec:overview}, highlighting their salient features. Results are presented in Section \ref{sec:results}, with a focus on the evolution of ensemble forecasts with lead time; this section is organized in subsections, addressing in sequence forecasts of storm occurrence, position, thermal and kinematic structure and intensity. The results are finally discussed in Section \ref{sec:conclusions}, alongside concluding remarks.

\section{Data and methods} \label{sec:methods}

In this section, an illustration is provided of the methods and techniques that are used to analyse the ECMWF operational analysis and ensemble forecast data in order to apply the object-based approach introduced in Section \ref{sec:intro}. Mean sea level pressure (MSLP) lows are first identified and tracked in both analysis and ensemble data. Forecast cyclones are then matched to the reference cyclone in the analysis, using a DTW technique to maximize the similarity of the trajectories. Forecasts are evaluated over a time interval as opposed to a fixed forecast time: the choice of the time interval depends on the cyclone's intensity as well as dynamical and thermal structure measured by suitable metrics. A short description is finally provided of the graphics used in evaluating the ensemble forecast statistics.

\subsection{Data} \label{subsec:data}

ECMWF operational analysis data is used as reference data to verify ensemble forecasts, which are initialized twice daily (at 0000 and 1200 UTC) and consist of 50 perturbed forecasts or members and a control forecast. Time resolution is 6 hours for both analysis and ensemble data.

Both the high-resolution, deterministic model (HRES), which is used to generate analysis data, and the ensemble prediction system (ENS) have undergone some changes during the time period considered in this study (2011-2017). Between 2011 and early 2016, horizontal grid spacing is 16 km for the HRES and 32 km for the ENS; afterwards, grid spacing decreases to 9 km for the HRES and 18 km for the ENS. Five and three events respectively occurred during these two time periods (see also Table \ref{tab:features} in Section \ref{sec:overview}). Vertical resolution also changed for both HRES and ENS during the analysed time period. The reader is referred to the ECMWF webpage for a detailed documentation of model changes and updates: \url{https://www.ecmwf.int/en/forecasts/documentation-and-support/changes-ecmwf-model}.

\subsection{Cyclone detection and tracking} \label{subsec:cyclone_tracking}

Many available cyclone detection methods \citep{neu_2013} are not suitable for Medicanes, which have a much smaller radius compared to most types of cyclones \citep[see e.g.][]{miglietta_2011,picornell_2014}. This issue is especially apparent when the input data has a relatively low horizontal resolution \citep{walsh_2014} which is the case for ECMWF ensemble forecast data. For this reason, we have developed a new detection method to identify pressure lows in both analysis and forecast data. This method has proven effective in detecting very small cyclones, while still being capable of detecting larger cyclones as well as filtering out spurious ones produced by noise or orographic effects.

The cyclone detection method used in this study is based on MSLP contours, spaced at 1 hPa intervals, and low-level vorticity, defined as relative vorticity averaged over the 1000, 925 and 850 hPa levels. Given our focus on the mature phase of cyclones, only closed contours are considered, thereby neglecting open systems \citep[e.g. diminutive waves, see][]{hewson_2009}. Pressure lows are identified as objects falling into at least one of two categories: (1) a set of four or more concentric contours; and (2) a set of two or more concentric contours with a radial MSLP gradient of 5 hPa/400 km or larger, calculated within a 400 km distance from the centre of the innermost contour, over at least 4 consecutive 30\degree-spaced azimuthal directions. The second category is necessary to include also earlier stages of a cyclone in which its closed circulation is still developing but a small pressure low is already present at the boundary of a larger region of low pressure, with a large MSLP gradient in its vicinity.

Detected lows are discarded when at least one of the three following conditions is met: (1) the area of all MSLP contours exceeds 500000 square km (low is too large); (2) the area of the second innermost contour is 50 times or more larger than the area of the innermost contour and MSLP gradient is smaller than 5 hPa/400 km in all directions (low is considered noise); and (3) contours are too thin and irregularly shaped (low is considered noise – this typically occurs in the vicinity of high orography). The centre of each pressure low is finally placed where low-level vorticity reaches a local maximum within a 100 km distance from the MSLP minimum. The values of thresholds and parameters have been chosen conservatively, so as to minimise the number of discarded lows. The outcome of cyclone detection shows little sensitivity to small variations of these thresholds and parameters.

After being detected, pressure lows are tracked in time using a method adapted from \citet{hewson_2010}, which uses 1000-500 hPa geopotential height difference (thickness) and 500 hPa wind speed: while a short description is given here, the reader is referred to the article above for a detailed explanation. In this tracking scheme, a likelihood score (expressed in km) is computed for each possible pairing of a pressure low at the previous output time and one at the current output time (hereafter referred to as ``past low'' and ``present low'', respectively). The score estimates the likeliness of the pairing being correct: that is, how likely the present low is the result of the past low advancing to a new position.

In the present study, the likelihood score is built on two parameters: half-time separation and thickness change. Half-time separation is the distance between the past and the present low, after they are moved forward and backward in time, respectively, for 60\% of the time interval, considering 500 hPa wind as the steering flow. Thickness change is the difference in 1000-500 hPa thickness between the positions of the past and the present low. A third parameter which was originally used in the likelihood score formula, namely feature type transition \citep{hewson_2010}, is kept fixed to 60\% \citep[][Table 2]{hewson_2009} when calculating the likelihood score, as the only type of feature considered in the present study is the closed low.

The smaller the likelihood score is, the more likely the pairing is correct – a low score results from a small half-time separation and a small thickness change. Pairings are discarded if the past and the present low are more than 600 km apart or if their likelihood score is higher than 700 km. After computing the likelihood score for all possible pairings, they get ranked from the lowest (most likely) to the highest (least likely). The ranking is finally read from top to bottom and each pairing is either accepted, if neither low was already previously paired, or rejected otherwise. When a pairing is accepted, the present low becomes the last element of the track that contains the past low. At the end, the remaining present lows form new tracks.

\subsection{Evaluation metrics} \label{subsec:metrics}

In order to evaluate ensemble forecasts four metrics are used that are deemed to provide an adequate picture of each cyclone's intensity, kinematics and thermal structure: these are MSLP, symmetry, compactness and the upper-level thermal wind. The statistics of ensemble forecasts of these metrics will be examined in Section \ref{sec:results}, together with those of storm position forecasts (see also explanation in Subsection \ref{subsec:eval_forecasts}).

Storm intensity is represented by the cyclone's lowest MSLP. The intensity of Medicanes may be slightly underestimated by ECMWF operational analysis data due to insufficient horizontal resolution, an effect that is estimated to be around 2 hPa \citep[see e.g.][]{cioni_2016,pytharoulis_2018}. An even larger underestimation can be expected for ensemble forecasts, given their resolution \citep{picornell_2014,walsh_2014} which is half of that of the analysis data.

In order to quantify the symmetry of the cyclone's low-level circulation, a symmetry parameter $S$ is defined for any MSLP contour as follows: $S=1+\arctan(\pi(4\pi A/P^2-1))$ where $A$ is the area and $P$ the perimeter of the contour. This seemingly complex formula is based on a straightforward expression of symmetry ($A/P^2$); this function is then scaled so that maximum symmetry – a perfectly round contour – equals 1 ($4\pi A/P^2$) and finally stretched by applying the arctangent, so that values of high symmetry are more widely spaced (they would otherwise tend to bunch towards 1). The last step allows to better identify the highly symmetric, mature phase of the cyclone and interpret the ensemble statistics more clearly. The $S$ parameter attains values of approximately 0.5, 0.2, 0 for ellipse-shaped contours having a minor axis of length 1 and a major axis of length 2, 3, 5 respectively. The symmetry parameter of any pressure low (hereafter referred to as just ``symmetry'', for the sake of brevity) is obtained by averaging S over the four innermost MSLP contours, spaced at 1 hPa intervals. The $S$ parameter constitutes a representative metric for the high symmetry attained by Medicanes during their mature, tropical-like phase, with $S$ values exceeding 0.8, as opposed to their early stages as well as the majority of extratropical cyclones which have much lower $S$ values.

Medicanes also tend to be much smaller than extratropical cyclones, as already pointed out, with strong pressure gradients in the vicinity of their centres. In order to give a measure of such gradient, a ``compactness'' parameter (hereafter referred to as just ``compactness'') is defined as the azimuthally averaged radial MSLP gradient within a 150 km radius around the cyclone centre, expressed in hPa/100 km.

Finally, the $-V_T^U$ metric is chosen to quantify the cyclone's upper-level thermal structure, namely its cold or warm core. The $-V_T^U$ parameter represents upper-level thermal wind, being one of three quantities defining the cyclone phase space (CPS) introduced by \citet{hart_2003}. A positive (negative) sign of $-V_T^U$ indicates an upper-level warm (cold) core, while the absolute value is proportional to its magnitude. Given the lower height of the tropopause in the midlatitudes with respect to the tropics \citep{picornell_2014} and the smaller size of Medicanes compared to tropical cyclones \citep{miglietta_2013}, $-V_T^U$ is calculated in a slightly different way from \citet{hart_2003}, using a smaller radius of 100 km and lower levels of 925, 700 and 400 hPa similarly to \citep{picornell_2014}. A 12-hour running mean is used in the present study to smooth the CPS trajectories, differently from \citet{hart_2003} who uses a 24-hour mean: this choice is motivated by the short life of most Medicanes and of their tropical-like phase in particular.

\subsection{Choice of the evaluation time interval} \label{subsec:ETI}

Ensemble forecasts are evaluated against analysis data over a time interval rather than at a single forecast time. This approach has the benefit of enhancing signals, in that it can spot desired features – e.g. a storm intensity maximum – over a larger set of forecast times, thereby overlooking small timing errors (e.g. the maximum occurring a few hours earlier or later than forecast). The rationale for our approach is to focus on specific storm features and consider a forecast to be sufficiently accurate if the features are successfully predicted, albeit at a slightly incorrect time. This strategy is valuable especially in extracting information from early forecasts.

The evaluation time interval (hereafter referred to as ETI) is 24 hour long, corresponding to 5 points (i.e. 5 output times) of the cyclone track extracted from analysis data (hereafter ``reference track''). Slightly shorter or longer ETIs were tested before settling on 24 hours, showing little sensitivity. The ETI is subjectively selected to best represent the mature, tropical-like phase of the cyclone, on the basis of the symmetry, compactness and $-V_T^U$ parameters introduced in Subsection \ref{subsec:metrics} (MSLP is not used as the mature phase of Medicanes often does not correspond to their most intense one).

\begin{figure}[h]
\centering
\includegraphics[width=0.8\textwidth]{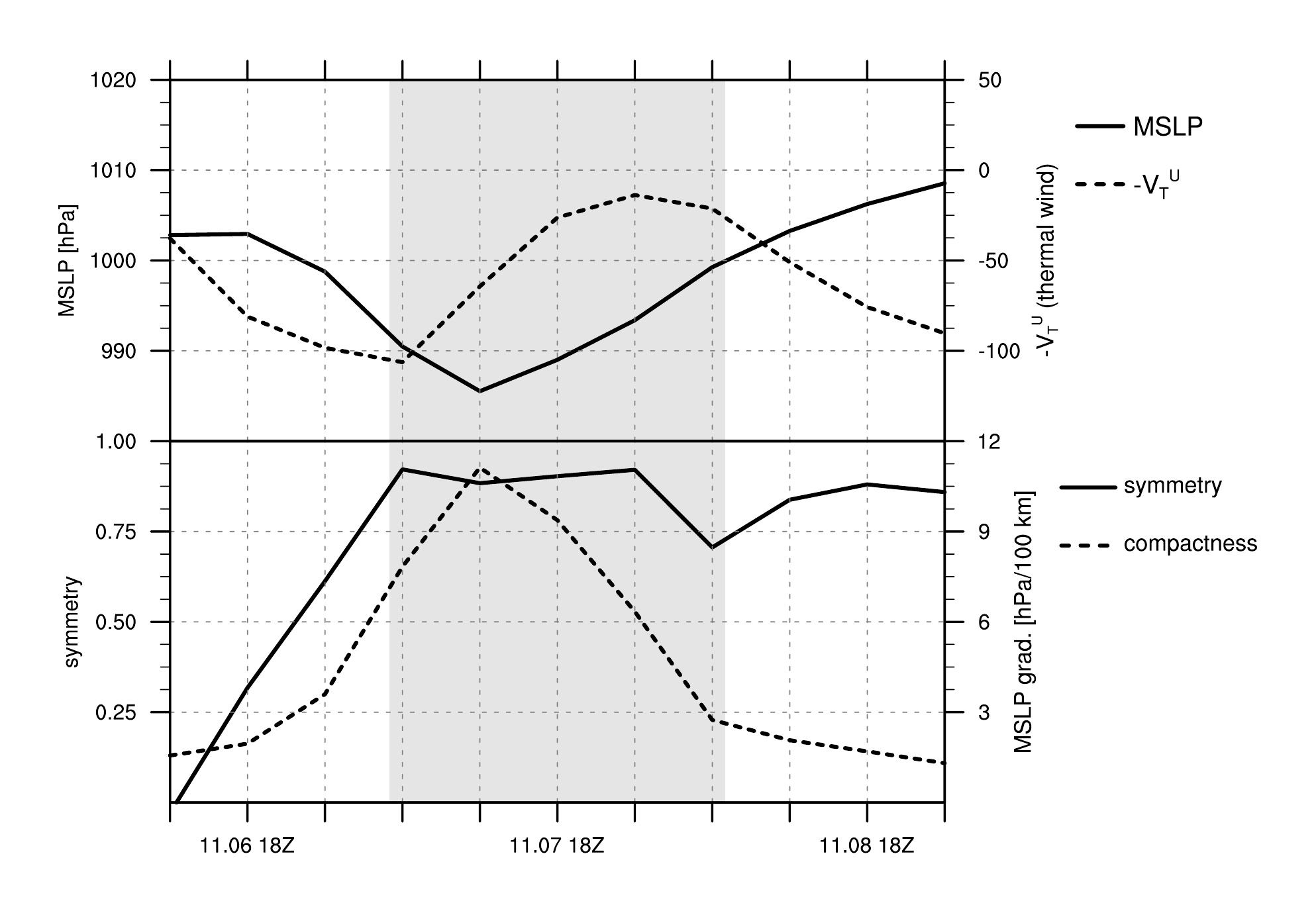}
\caption{Features of storm Qendresa (November 2014) as retrieved from analysis data. Upper panel: MSLP (hPa) and upper-level thermal wind $-V_T^U$; lower panel: symmetry and compactness. The ETI is highlighted in gray: in this case the selected period is from 0600 UTC on 7 November 2014 to 0600 UTC on 8 November 2014.}
\label{fig:qendresa_features}
\end{figure}

An example of ETI selection is shown in Figure \ref{fig:qendresa_features}. As a first step, the 5 consecutive reference track points having the highest average $-V_T^U$ value are selected, given that $-V_T^U$ is considered the most relevant parameter in distinguishing tropical-like cyclones from fully baroclinic cyclones \citep[see e.g.][]{mazza_2017}. As a second step, the initial 5-point selection is shifted by at most 2 points, corresponding to maximum 12 hours earlier or later. This adjustment is only applied when necessary, to select output times with as high symmetry and compactness as possible: for storm Qendresa (Figure \ref{fig:qendresa_features}), for instance, the initial selection is shifted 1 point to the left (6 hours earlier) thereby increasing the average value of symmetry and compactness.

\subsection{Track matching} \label{subsec:track_matching}

The tracking procedure outlined in Subsection \ref{subsec:cyclone_tracking} is used for each storm to retrieve the reference track as well as tracks of MSLP lows in individual forecasts. The next step is then to compare all tracks from a single member of the ensemble with the reference track in order to find the best match. In order to avoid penalizing (small) discrepancies in the timing of storm motion and take into account the overall spatio-temporal similarity between tracks, we use a dynamic time warping (DTW) technique \citep{berndt_1994} which has been successfully applied to a recent case study of North Atlantic tropical transition \citep{michael_chris}. Originally developed for speech recognition \citep{sakoe_1978}, DTW is able to match two time series nonlinearly, thereby taking into account differences in signal speed and providing a more intuitive matching \citep{keogh_2005}. Using the DTW technique to match cyclone tracks allows to focus on the spatial accuracy of forecasts, ignoring small (local) timing errors as long as the forecast track bears a high spatial similarity to the reference one. The average time difference between DTW-paired track points may be later used to assess whether there is an early or late bias. In the following, we briefly describe the structure of the DTW technique to illustrate how it is applied to matching cyclone tracks. The reader is referred to \citet{berndt_1994} and \citet{keogh_2005} for more detailed explanations of the algorithm.

\begin{figure}[h]
\centering
\includegraphics[width=\textwidth]{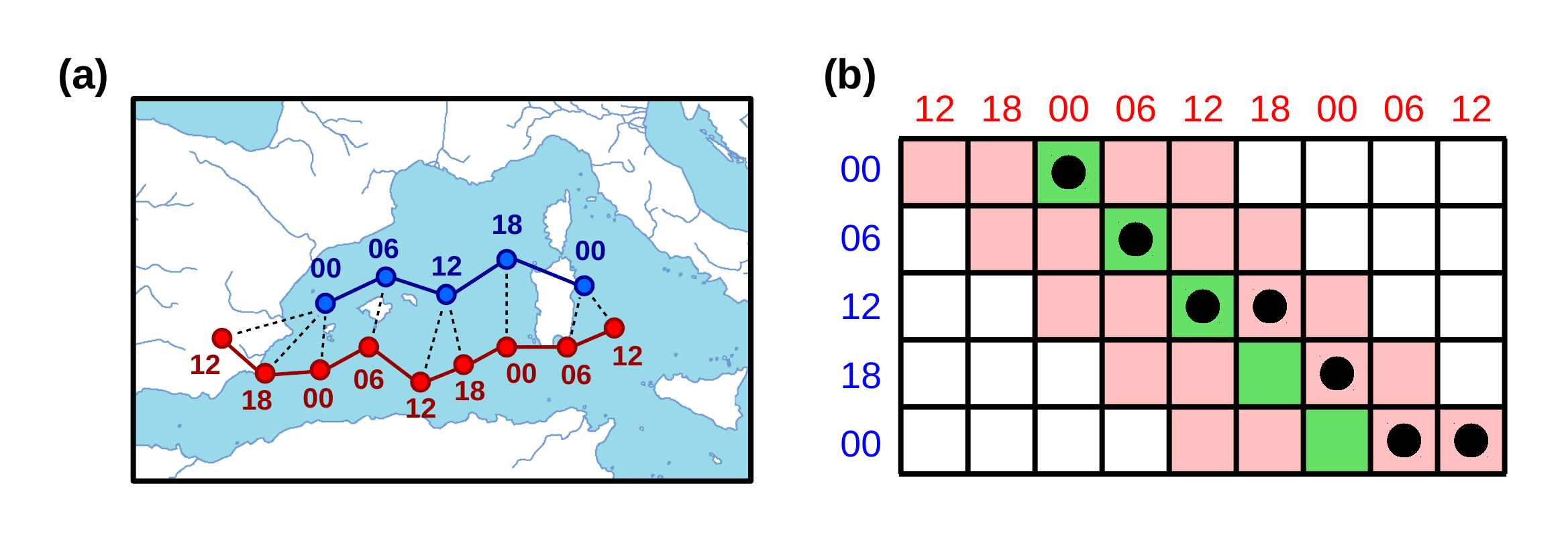}
\caption{Example of DTW matching of the reference track (blue) and a forecast track (red). Numbers denote UTC times. a) Spatial match; matched track points are highlighted by a black dashed line. b) Cumulative distance matrix $M$ and warping path, represented as black filled circles, for the track match in a). The warping window is highlighted in red while the equal-time match is highlighted in green.}
\label{fig:dtw_match}
\end{figure}

DTW requires first a suitable metric to express the spatial distance between each pair of track points. We choose great circle distance but note that any distance metric could be used in principle. The aim is then to minimize the overall distance between the two input tracks $R = r_1, r_2, \dots r_m$ and $S = s_1, s_2, \dots s_n$ by finding the best possible way of matching them. In order to do so, a $m \times n$ distance matrix $D$ is first computed: $D(i,j) = d(r_i,s_j)$ for each $i = 1, \dots m$ and $j = 1, \dots n$, where $d(r_i,s_j)$ is the spatial distance between the $r_i$ and $s_j$ track points. A cumulative distance matrix $M$ is then defined recursively as follows: $M(i,j) = D(i,j) + \min[D(i-1,j),D(i-1,j-1),D(i,j-1)]$. The best match is finally obtained as the \emph{warping path}, defined as the succession of $M$ elements minimizing the cumulative distance at every point. Each element of the warping path represents a pair of matched track points, as shown in the fictitious example in Figure \ref{fig:dtw_match}. We note that the two tracks have here different lengths and that multiple points of one track are matched to a single point of the other.

A DTW technique is usually applied with some constraints, which introduce physically meaningful requirements \citep{berndt_1994}. Monotonicity and continuity constraints are first imposed to assure that all track points are matched at least once and with increasing time. A warping window (highlighted in red in Figure \ref{fig:dtw_match}) only allows the warping path to exist in the vicinity of the diagonal of the $M$ matrix (i.e. the succession of equal-time elements), thereby restricting the time difference between any pair of matched track points to a maximum absolute value of 12 hours. Using a warping window ensures a physically meaningful track match, preventing the match of two points that are spatially close but very distant in time. Finally, boundary conditions require the warping path to start from (end at) the forecast track point that is closest to the first (last) analysis track point, to prevent the algorithm from matching too many far away forecast track points to the first or last analysis point, which it would be forced to do if the forecast cyclone moves fast (an example is seen in Figure \ref{fig:dtw_match}, where the first two forecast track points are not matched to the first analysis track point). These conditions ensure that similarity is maximised in the matching process.

DTW is applied to match the reference track's 24-hour ETI to each track in an ensemble member. Only forecast tracks that are at least 24 hours long (5 output times) are considered. Furthermore, a 48-hour interval is selected from each track, spanning the ETI plus further 12 hours (2 output times) at both ends, to meet the warping window constraint. If the forecast track only exists for a fraction of these 48 hours, only the existing part is considered (the example in Figure \ref{fig:dtw_match} shows the longest possible forecast track, at 48 hours or 9 output times). The spatio-temporal distance between (the selected interval of) the forecast and reference tracks is finally computed in two steps: the average distance between a single reference track point and all associated forecast track points is first computed; the final track distance is then obtained by averaging the result of the above calculation over all reference track points.

For a given ensemble member, forecast tracks having a 600 km or larger spatio-temporal distance from the reference track are discarded. This threshold has been chosen after testing the sensitivity of the results to its value, similarly to \citet{michael_chris}. If no forecast tracks are left, the member is considered to have no cyclone (such members will be named ``no-storm members'' hereafter). Otherwise, the track with the shortest spatio-temporal distance is considered to be the best match, i.e. the most similar to the reference track. Members having a best match will be named ``storm members'' hereafter.

\subsection{Evaluation of ensemble forecasts} \label{subsec:eval_forecasts}

Ensemble forecasts of the eight storms are evaluated in Section \ref{sec:results}. Given that forecasts are evaluated over a time interval rather than at a fixed forecast time, lead times refer to the central time of the ETI. For each storm, the latest forecast considered is the one initialized either at the beginning of the ETI, if this begins at 0000 or 1200 UTC, or 6 hours earlier, if the ETI begins at 0600 or 1800 UTC. For the sake of simplicity, in Section \ref{sec:results} plots displaying the evolution of forecast statistics with lead time, the latest forecast is always labeled as ``0.5 day'' (the small 6-hour difference introduced by varying beginning time of the ETIs does not affect the results).

Section \ref{sec:results} plots show the statistics of 16 consecutive ensemble forecasts of one of the metrics introduced in Subsection \ref{subsec:metrics}, for a maximum lead time of 8 days. Box-percentile plots \citep{esty_2003} are preferred to standard box-and-whisker plots in that they display the whole distribution of input data: the width of each irregular ``box'' is proportional to the percentile $p$ of the ordinate if $p \leq 50$, or to $100-p$ if $p > 50$; the maximum width is thus reached at the median, while outliers are revealed by thin spikes at each tail. These forecasts are relative to the extreme value of each metric (the lowest value for MSLP, the highest value for the others) computed for each storm member within the DTW-matched interval of the best track; the reference value shown in the plots is also the extreme value computed in the reference track's ETI. 

Storm position forecast statistics (Figure \ref{fig:position_error}) are investigated by means of EOF analysis. For each storm member, a 2D storm position error is expressed as the average longitude and latitude difference between the reference track and the forecast track, computed using the DTW-matched points in an analogous manner as the spatio-temporal distance. EOF analysis \citep{wilks_2011} is then performed on all 2D error values for each forecast (one value for each storm member). The eigenvectors of their covariance matrix define a rotated coordinate system where variability is maximized along the x-axis. The spread of storm position forecast errors is proportional to their variance in this coordinate system and it is represented in Figure \ref{fig:position_error} as an ellipse whose axes are aligned to the ones of the rotated system and have lengths proportional to the variance along each eigenvector. This compact representation of storm position errors provides an immediate visual grasp of their extent and spatial distribution.

\section{Overview of the storms} \label{sec:overview}

\begin{figure}[!h]
\centering
\includegraphics[width=\textwidth]{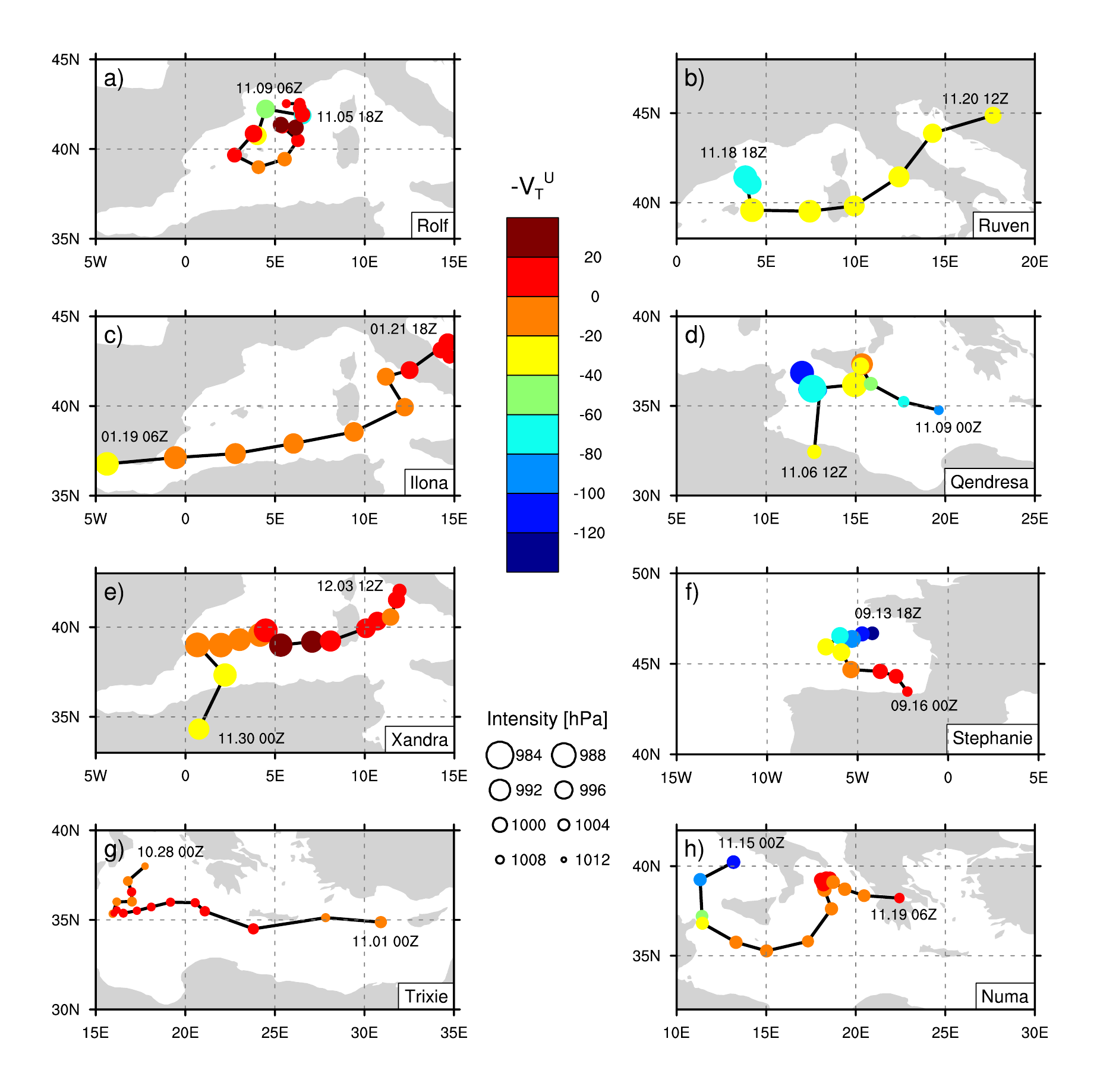}
\caption{Tracks of the eight storms. The colour of each circle represents the $-V_T^U$ value, while its size represents the MSLP value.}
\label{fig:all_storms}
\end{figure}

The eight storms analysed in this study are briefly illustrated here. A summary of their main features as retrieved from analysis data is given in Table \ref{tab:features}, where storm duration, period and region of occurrence are provided along with extreme intensity, symmetry, compactness, 10 m wind speed and upper-level thermal wind. Storm names were chosen by the Institute of Meteorology at the Free University of Berlin. The storm trajectories, intensity and upper-level thermal wind values are displayed in Figure \ref{fig:all_storms}.

\begin{table}[!h]
\centering
\caption{Period and region of occurrence, duration, MSLP [hPa], symmetry, compactness [hPa/100 km], 10 metre wind [m$\text{s}^{-1}$], upper-level thermal wind $-V_T^U$ for the eight storms, as inferred from operational analysis data. Values are the lowest (for MSLP) or highest (for other quantities) reached in the life cycle of the cyclone. 10 metre wind is computed in a 300 km around the centre of the storm.}
\label{tab:features}
\begin{adjustbox}{width=\textwidth}
{\small
\begin{tabular}{llllccccc}
\headrow
\thead{Storm} & \thead{Period} & \thead{Region} & \thead{Duration} & \thead{MSLP} & \thead{Symmetry} & \thead{Compactness} & \thead{10 m wind} & \thead{$\bm{-V_T^U}$} \\
Rolf & Nov. 2011 & WM & 96 h & 997 & 0.95 & 6.6 & 22 & 26 \\
Ruven & Nov. 2013 & WM, TS, AS & 48 h & 990 & 0.85 & 3.8 & 22 & -31 \\
Ilona & Jan. 2014 & WM, TS, AS & 60 h & 991 & 0.83 & 3.2 & 23 & 8 \\
Qendresa & Nov. 2014 & SM & 60 h & 986 & 0.92 & 10.8 & 27 & -14 \\
Xandra & Nov./Dec. 2014 & WM, TS & 84 h & 989 & 0.95 & 3.6 & 19 & 22 \\
Stephanie & Sep. 2016 & BB & 54 h & 998 & 0.96 & 6.0 & 22 & 11 \\
Trixie & Oct. 2016 & SM, EM & 96 h & 1005 & 0.96 & 4.9 & 24 & 18 \\
Numa & Nov. 2017 & TS, SM, IS & 120 h & 1002 & 0.98 & 5.1 & 19 & 20 \\
\hline  
\end{tabular}
}
\end{adjustbox}
\begin{tablenotes}
\item WM = Western Mediterranean; SM = Southern Mediterranean; EM = Eastern Mediterranean; TS = Tyrrhenian Sea; AS = Adriatic Sea; IS = Ionian Sea; BB = Bay of Biscay.
\end{tablenotes}
\end{table}

Of the eight storms, four developed or spent a significant part of their lifetime over the Western Mediterranean, three over the Southern Mediterranean and the Ionian Sea: these two regions are indeed hotspots for Medicanes \citep{cavicchia_2014_clim}. In contrast, storm Stephanie is technically not a Medicane, in that it occurred outside of the Mediterranean Sea. Stephanie has been included in our study as it exhibited the same tropical-like traits as Medicanes \citep{biscane}, formed under similar large-scale circulation patterns and occurred in a region that is geographically and climatologically close to the Mediterranean. Five of the eight storms occurred in November, the most frequent month in our sample; the three remaining storms occurred in September, October and January, respectively. We note in passing that this temporal distribution differs from the one relative to the 1948-2011 climatology produced by \citet{cavicchia_2014_clim}, which has a maximum in January.

The data in Table \ref{tab:features} and the cyclone tracks in Figure \ref{fig:all_storms} show the high heterogeneity of the eight storms in terms of their duration (Ruven developed rapidly and only lasted 48 hours, Numa remained almost statically over the Ionian Sea for 36 hours and lasted 120 hours), intensity (almost 20 hPa difference between the most intense, Qendresa at 986 hPa, and the least intense, Trixie at 1005 hPa), compactness (Qendresa reached 10.8 hPa/100 km MSLP gradient, Ilona only 3.2 hPa/100 km) and thermal structure (most storms developed a moderate upper-level warm core, yet storms Ruven and Qendresa failed to attain one, with upper-level thermal wind $-V_T^U$ peaking slightly short of zero). We observe here that even though storm Qendresa did not attain an upper-level warm core, it is widely recognised as a Medicane \citep{pytharoulis_al_2017,pytharoulis_2018,cioni_2018}.
In fact, a unique, objective definition of what constitutes a Medicane has not yet been established in the literature \citep{fita_2018}. However, all storms analysed in this study share some distinctive traits, in that at some point during their life cycle they shrink notably, acquiring a highly axisymmetric circulation with strong MSLP gradients while quickly moving towards positive values of upper-level thermal wind (i.e. building a warm core). Another feature shared by most storms is their weakening or even fading after making landfall, which is consistent with the fact that Medicanes, similarly to tropical cyclones, are strongly influenced by surface fluxes \citep{fita_2007,tous_fluxes}.

\section{Results} \label{sec:results}

In this section, forecasts of several metrics are analysed, with a focus on the evolution of ensemble statistics with lead time (hereafter referred to as LT). Forecasts of cyclone occurrence are first examined in Subsection \ref{subsec:results_occurrence}. Cyclone position forecasts are then explored in Subsection \ref{subsec:results_position}. To analyse the storms' thermal structure, upper-level thermal wind forecasts are examined in Subsection \ref{subsec:results_thermal}. The kinematic structure and intensity of the eight storms are finally discussed in Subsection \ref{subsec:results_kinematic}. 

\begin{figure}[!h]
\centering
\includegraphics[width=\textwidth]{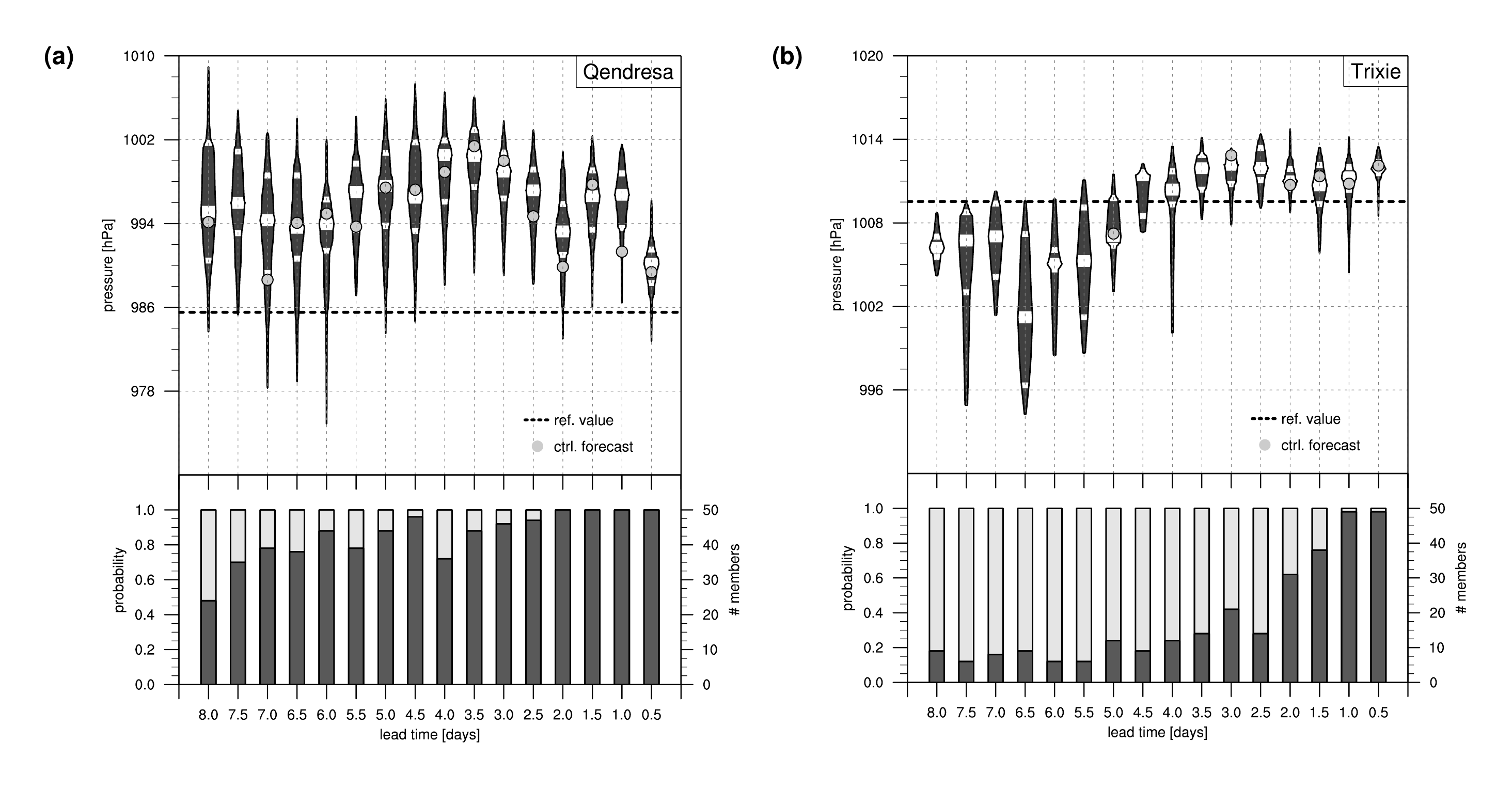}
\caption{Ensemble forecasts of MSLP for storms Qendresa (left) and Trixie (right). Upper panels: box-percentile plots, with white stripes marking the 25th, 50th (median) and 75th percentiles; the yellow circle shows the control forecast value (provided it has the cyclone); the dashed line is the operational analysis value. Lower panels: number of members having the cyclone (no cyclone) are represented by blue (gray) bars.}
\label{fig:results_intro}
\end{figure}

Two examples of the evolution of ensemble forecasts are provided in Figure \ref{fig:results_intro}, which shows MSLP forecasts for storms Qendresa and Trixie. These two cases illustrate the high variability among both Medicane features (see also Section \ref{sec:overview}) and their forecasts. Qendresa is the deepest cyclone among the eight cases, with 986 hPa minimum pressure in the ETI. For this storm, the probability of cyclone occurrence (i.e. the number of storm members) is already high at 7.5 days LT (Figure \ref{fig:results_intro}a) and remains high throughout. Conversely, storm intensity is consistently underpredicted, with the ensemble median up to 14 hPa higher than the analysis value. A small but evident dip is seen around 4 days LT for both occurrence probability and storm intensity. On the other hand, Trixie is the weakest cyclone in our list, with over 1009 hPa, albeit very long-lived, with a lifetime of 96 hours (Table \ref{tab:features}). For this storm, occurrence probability is much lower than 0.5 at LTs longer than 3 days, with a considerable increase between 2.5 and 1 day LT (Figure \ref{fig:results_intro}b). The intensity of Trixie is first overpredicted in early forecasts up to 5 days LT, then slightly underpredicted with a greatly reduced spread. 

In both these cases, the evolution of ensemble forecasts with lead time is far from gradual, with storm intensity forecasts showing little convergence towards the analysis value for Qendresa while an early convergence is followed by a plateau for Trixie; the probability of cyclone occurrence is consistently high for Qendresa, whereas it is very low for early forecasts, but grows rapidly for late forecasts for Trixie: we name such rapid increases \emph{forecast jumps} as already noted in Section \ref{sec:intro}.

\subsection{Cyclone occurrence forecasts} \label{subsec:results_occurrence}

Medicanes develop because of a combination of factors spanning multiple spatial and temporal scales and are therefore low frequency events \citep{cavicchia_2014_clim}. Early signals of the occurrence of a cyclone, as seen in ensemble forecasts 5-8 days in advance, are then to be considered a first valuable piece of prognostic information. For this reason, cyclone occurrence forecasts are examined as a first insight into the predictability of the eight storms.

\begin{figure}[!h]
\centering
\includegraphics[width=0.8\textwidth]{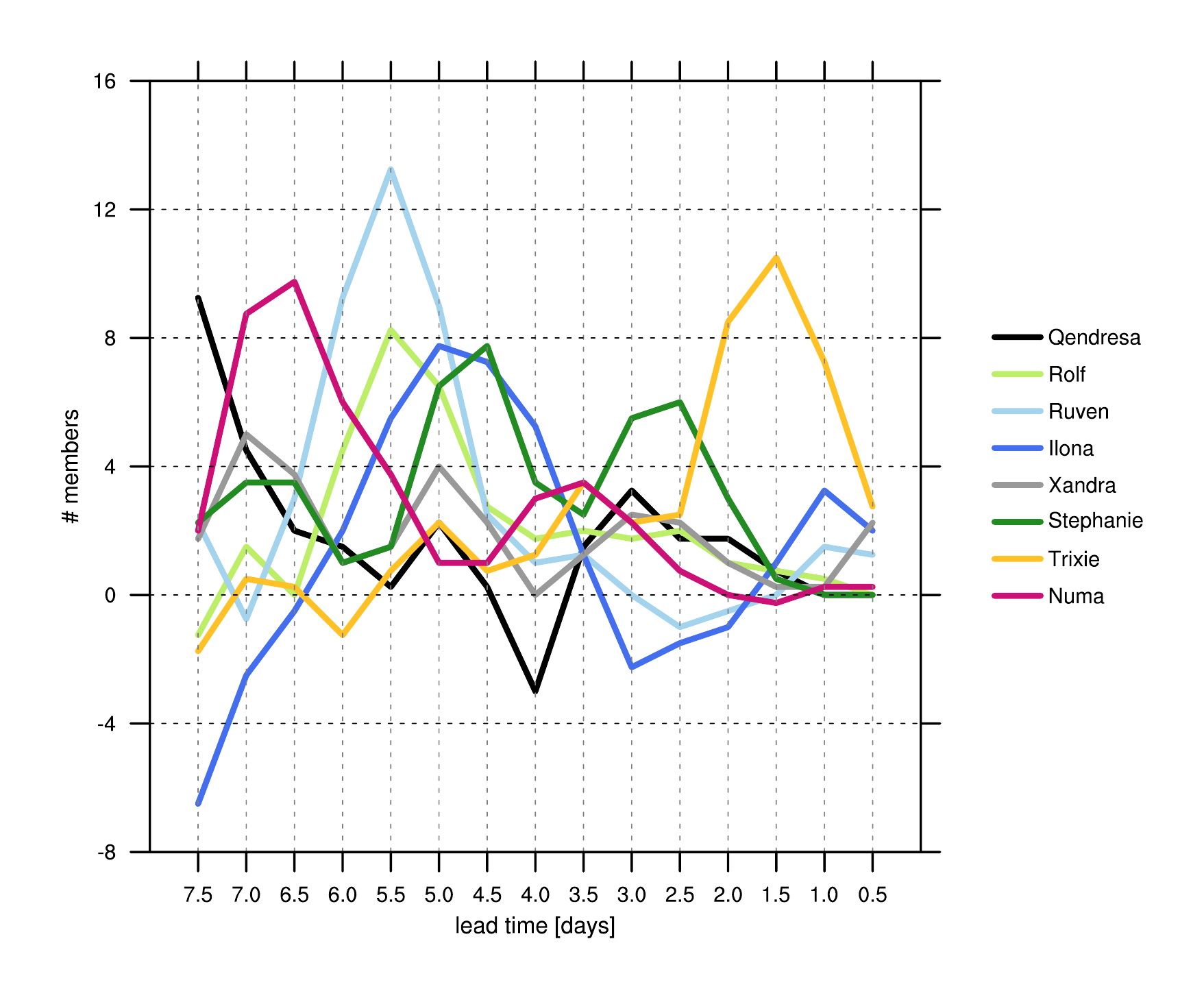}
\caption{Difference in the number of storm members (proportional to cyclone occurrence probability) between current and previous forecast, for each lead time, for all cases. Values are smoothed with a 1-2-1 running mean to reduce noise.}
\label{fig:diff_plot}
\end{figure}

It is reasonable to expect a gradual increase with lead time of the probability of cyclone occurrence. This is not the case for most storms analysed in the present study, as forecasts often exhibit a distinctly rapid increase in occurrence probability at some lead times (forecast jump). In order to extract such signals, the difference in the number of storm members between consecutive forecasts is computed and shown in Figure \ref{fig:diff_plot}. Seven out of eight cases exhibit distinct positive peaks, i.e. a rapid increase in occurrence probability over a short interval of lead time: Qendresa (7.5 days LT), Numa (around 7 days), Ruven (6 to 5 days), Rolf (around 5.5 days), Ilona (around 5 days), Stephanie (double peak around 5 and 3 days) and Trixie (2 to 1 day). These peaks stand out above the bulk of values which are contained between -1 and 5. Only one case, Xandra, shows a gradual increase of occurrence probability throughout all forecasts.

We note here that six out of seven occurrence forecast jumps are found at lead times longer than 4 days. A notable exception is Trixie, for which occurrence probability does not increase above 50\% until 2 days LT. These results are consistent with the low atmospheric likelihood of Medicane occurrence and compatible with the hypothesis that occurrence probability increases significantly only when the forecast model's initial data contain sufficient information on all processes impacting Medicane formation at all scales. This may constitute an instance of predictability barrier, possibly similar to the spring predictability barrier known to impact ENSO predictions \citep{duan_2013,levine_2015} albeit arising at different temporal scales.

\subsection{Cyclone position forecasts} \label{subsec:results_position}

The impacts of Medicanes can be considerable \citep{cavicchia_2014_clim} but spatially limited to small regions, due to their small size. For this reason, an accurate prediction of their trajectory is key in preventing and mitigating the damages they produce locally. The next step in evaluating ensemble forecasts of the eight storms is then to examine their predicted position during their mature, tropical-like phase. 

\begin{figure}[!h]
\centering
\includegraphics[width=0.85\textwidth]{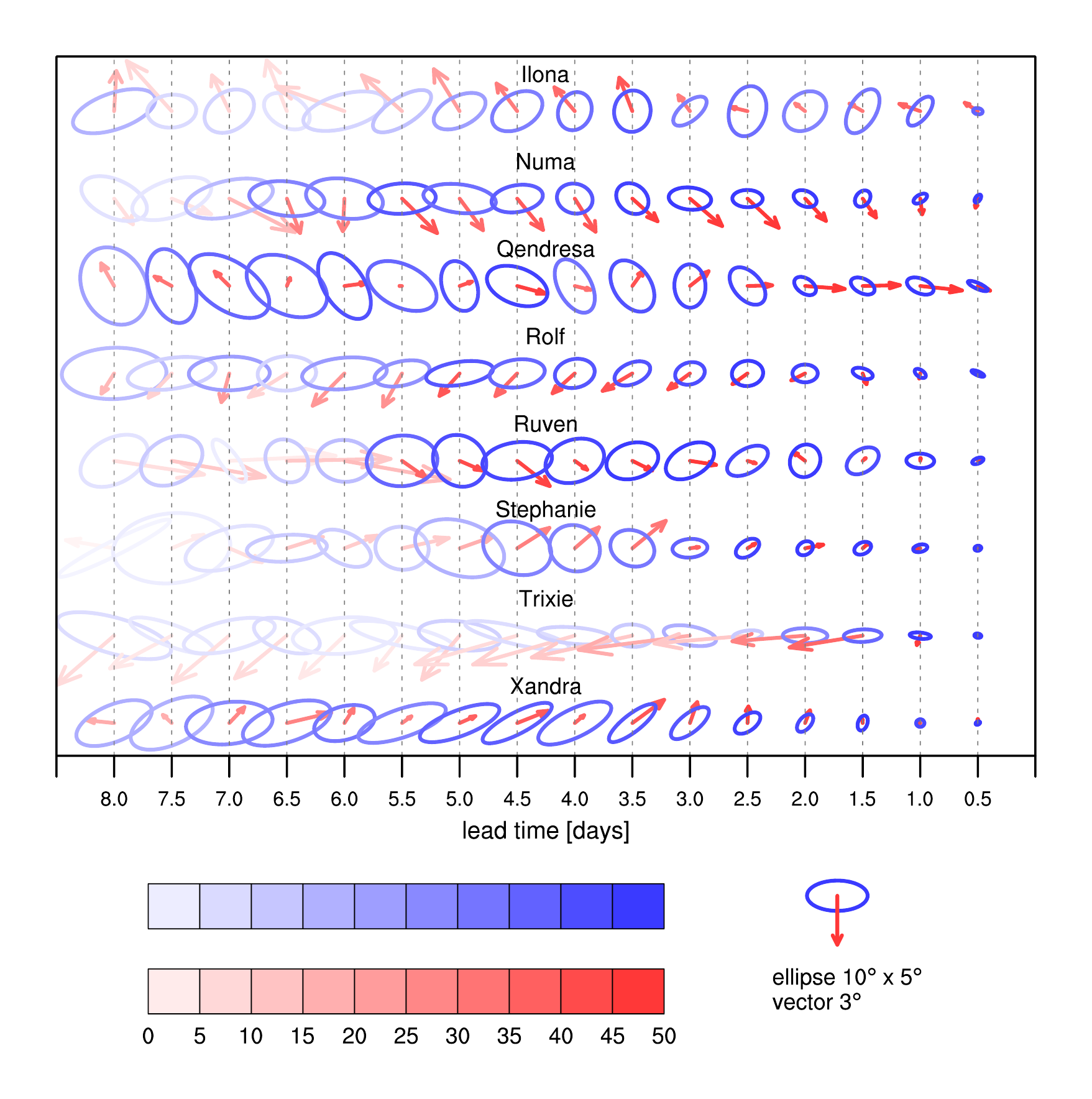}
\caption{Statistics of cyclone position forecasts. For any given forecast, only storm members are considered. The red arrow represents the median of position errors, its components being longitude (horizontal) and latitude (vertical). The blue ellipse is a bivariate normal distribution fit to the position errors, representing their spread; it is scaled so as to encircle 95\% of error points. The ellipse is oriented along the direction of maximum variability of the error values and its axes are proportional to their variance in the 2D rotated coordinate system defined by the eigenvectors of their covariance matrix (see Subsection \ref{subsec:eval_forecasts}). The more storm members, the brighter the colour of both ellipses and arrows.}
\label{fig:position_error}
\end{figure}

Cyclone position forecast statistics are shown in Figure \ref{fig:position_error}, where the median of position errors is represented as an arrow and forecast spread as an ellipse whose axes (and hence its area) are proportional to the variance of position errors (see Subsection \ref{subsec:eval_forecasts}). These forecasts appear to converge more gradually compared to cyclone occurrence forecasts, as demonstrated by the overall slow variation of the size and tilt of both arrows and ellipses over lead time\footnote{The convergence is towards a median and spread value which is very close to zero, but not exactly zero as forecasts are evaluated in a time interval.}. However, rapid changes of one or more of these quantities (referred to as ``position forecast jumps'') are visible at some lead times for some cases. For instance, a sudden decrease in spread, with the ellipse decreasing in size by 30\% or more between consecutive forecasts, is seen for Ilona (3 days LT), Numa (5.5 and 4.5 days), Qendresa (2 days), Rolf (5.5 days), Stephanie (3 days) and Xandra (6 days). Rapid changes in the magnitude of the median error are also apparent, e.g. for Ilona (3 days LT), Qendresa (5.5 days), Ruven (5.5 and 4 days), Stephanie (3 days), Trixie (1 day) and Xandra (3.5 days). Position forecast jumps occur in most cases at slightly shorter lead times than occurrence forecast jumps (difference is 2 days or less for Qendresa, Ilona, Numa, Ruven, Trixie, while the two jumps occur at the same lead time for Rolf and Stephanie). This suggests the existence of a causal link between increased occurrence probability and higher accuracy of position forecasts. 

It is worth noting that the spatial distribution of position errors tends to evolve slowly with lead time. For instance, forecasts exhibit a consistent northwestern bias for Ilona (i.e. the storm is predicted to occur too far to the northwest), a southern to southeastern bias for Numa, a southwestern one for Rolf, a northeastern one for Stephanie (at least until 3 days LT) and a large western to southwestern one for Trixie (although in this case with low occurrence probability until 2 days LT). Similarly, position errors are consistently distributed from west to east for Numa and Trixie, from NW to SE for Qendresa and from SW to NE for Rolf and Xandra. In summary, the region where the cyclone is predicted to occur often tends to remain the same between consecutive forecasts. This implies that early forecasts may already contain valuable prognostic information, in that the actual cyclone position may be approximately estimated early on by examining the spatial distribution of position forecasts. One explanation to this may be that certain areas of the Mediterranean Sea are more conducive to Medicane development than others \citep{tous_environments,cavicchia_2014_clim}, so that it is more likely that the cyclone is predicted to spend its mature phase in these regions. 

\subsection{Thermal structure forecasts} \label{subsec:results_thermal}

\begin{figure}[!hp]
\centering
\includegraphics[width=0.9\textwidth]{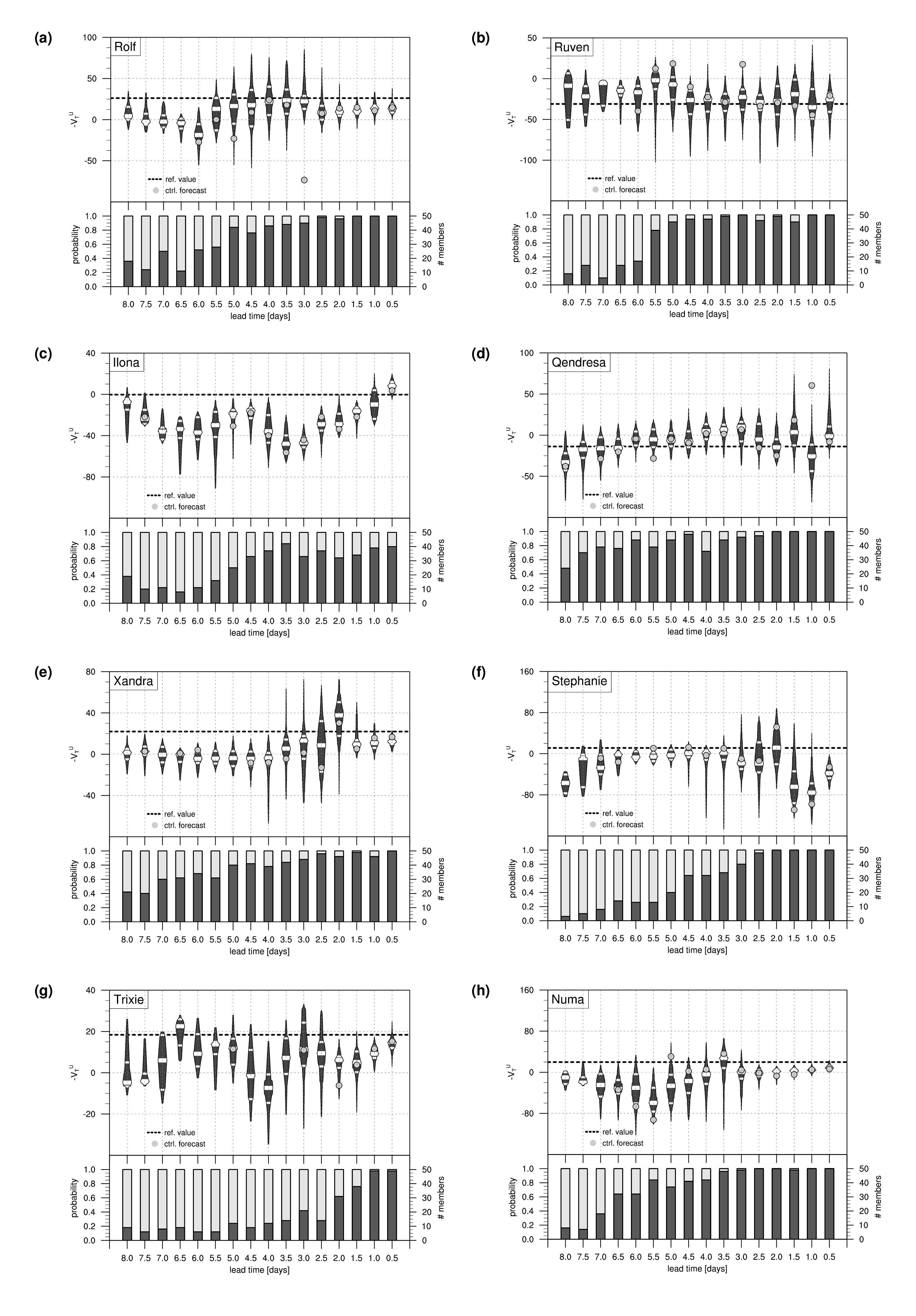}
\caption{As in Figure \ref{fig:results_intro}, but for upper-level thermal wind ($-V_T^U$) forecasts and for all storms.}
\label{fig:VTU_forecasts}
\end{figure}

After assessing whether a cyclone is going to occur or not and where it is going to occur, the next step is analysing its thermal structure. For this reason, we now examine forecasts of upper-level thermal wind, represented by the $-V_T^U$ parameter, which are shown in Figure \ref{fig:VTU_forecasts}. The evolution of these forecasts with lead time is generally neither gradual nor monotonic, as already noted with regard to forecasts of cyclone occurrence (Subsection \ref{subsec:results_occurrence}). Overall, the forecast spread does not consistently reduce with decreasing lead time, with some cases exhibiting a smaller (Qendresa and Stephanie) or comparable (Ilona, Rolf, Ruven, Xandra) spread at long lead times compared to the latest forecasts. Similarly, in some cases the median increasingly deviates from the analysis value with decreasing lead time, only to get closer again in later forecasts (e.g. Ilona, Numa, Qendresa, Rolf, Stephanie). 

Storms Rolf and Numa (Figure \ref{fig:VTU_forecasts}a and \ref{fig:VTU_forecasts}h, respectively) show a similar evolution, with the forecast median $-V_T^U$ increasingly drifting away from the analysis value and the spread increasing in parallel, until the median reaches a minimum and the forecast distribution is entirely below the analysis value (i.e. upper-level thermal wind is underpredicted). The forecast median then converges again towards the analysis, while the spread first decreases slowly, then much faster to eventually level off at short lead times. Storms Ilona and Ruven exhibit instead a contrasting evolution. For Ilona (Figure \ref{fig:VTU_forecasts}c) the forecast median drifts away twice from the analysis value with decreasing lead time, to eventually approach it in the latest forecasts; the spread oscillates considerably between consecutive forecasts throughout the period considered. For Ruven (Figure \ref{fig:VTU_forecasts}b) the median remains always somewhat close to the analysis value and the spread does not change considerably throughout the period considered.

A peculiar evolution is exhibited by storm Xandra (Figure \ref{fig:VTU_forecasts}e). Early forecasts consistently underpredict upper-level thermal wind, with very little spread. The spread then increases considerably between 4 and 2 days LT while the median $-V_T^U$ increases slightly. The spread finally decreases again rapidly in the latest forecasts while the median $-V_T^U$ remains slightly below the analysis value. We interpret this behaviour as follows: with little spread at the longer LTs, ensemble forecasts indicate with high probability the development of a weaker warm core or a cold core. The increase in spread with decreasing LT indicates that new information available in the initial conditions allows the development of a warmer upper-level core to occur in some ensemble members. The increase in spread then signifies the increase in probability of the Medicane to actually occur.

Forecasts of cyclone thermal structure do not appear to be consistently linked to occurrence probability. However, some cases show interesting behaviours: for instance, for Rolf (Figure \ref{fig:VTU_forecasts}a) the forecast median $-V_T^U$ approaches closely the analysis value only when probability is higher than 0.8; for Stephanie (Figure \ref{fig:VTU_forecasts}f) the increase in occurrence probability around 4.5 days LT appears to be associated at first to a broadening of the $-V_T^U$ forecast distribution and later to its shift towards lower values; for Trixie (Figure \ref{fig:VTU_forecasts}g) the rapid increase in occurrence probability at 2 days LT is associated to a reduction in the $-V_T^U$ forecast spread. 

In all cases, forecasts initialized when the cyclone has already developed have a much lower spread of upper-level thermal wind than previous forecasts and their median $-V_T^U$ also tends to be closer to the analysis value. This is probably explained by the inherently low probability of Medicane occurrence and the fact that the development of a warm core depends on a variety of factors, including small-scale ones such as surface fluxes, for which reason a preexisting cyclone constitutes a marked improvement in the initial conditions. We observe that in most cases the latest $-V_T^U$ forecast is more accurate than earlier ones, in terms of the median $-V_T^U$ being closer to the analysis value and the spread being lower. For this forecast, the analysis value lies within the ensemble distribution in all cases. Overall, this is evidence that the ECMWF ensemble model can adequately reproduce warm-core cyclones despite its relatively low horizontal resolution.

\subsection{Kinematic structure and intensity forecasts} \label{subsec:results_kinematic}

\begin{figure}[!h]
\centering
\includegraphics[width=\textwidth]{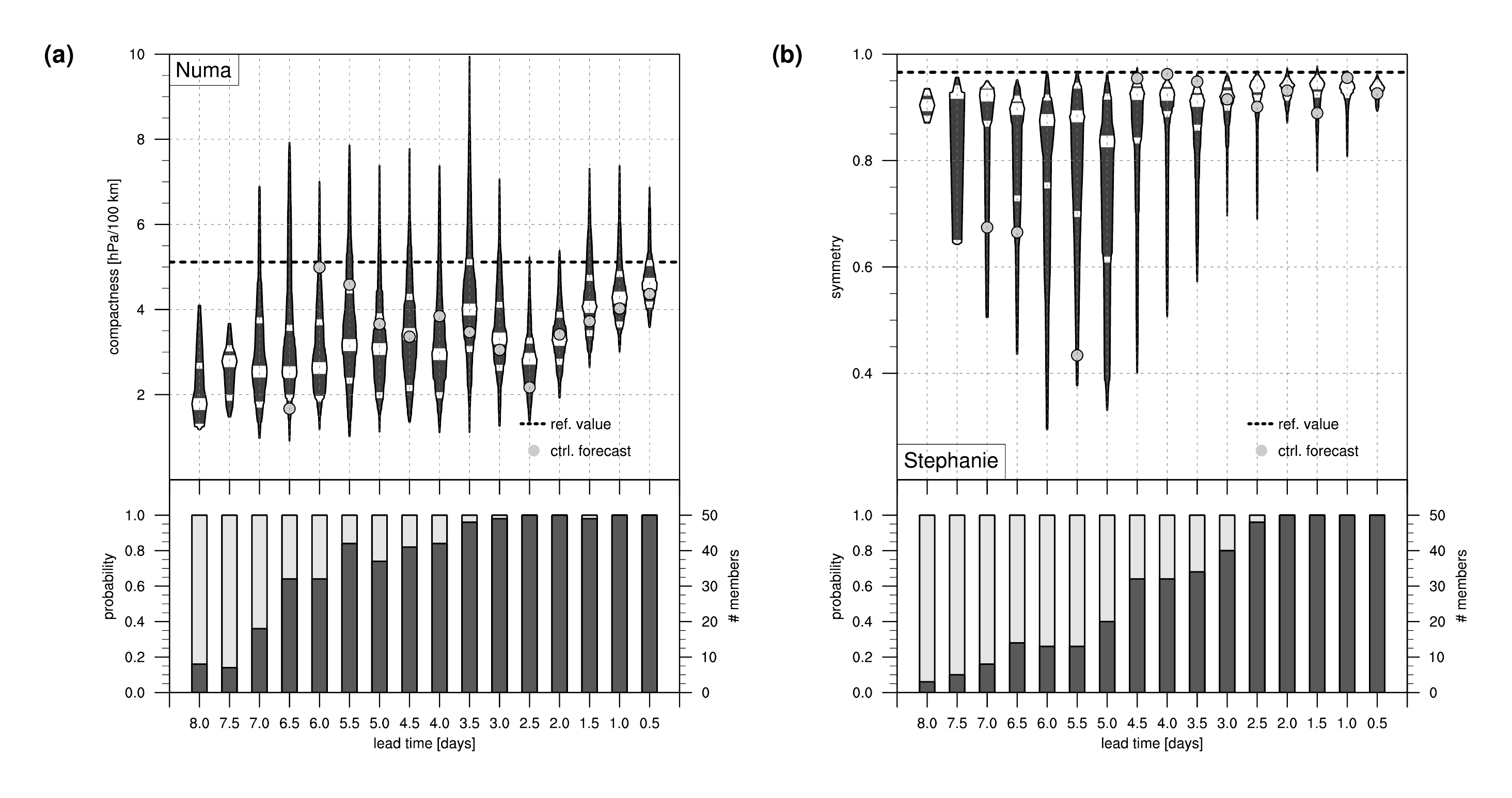}
\caption{As in Figure \ref{fig:results_intro}, but for a) the compactness forecast for Numa and b) the symmetry forecast for Stephanie.}
\label{fig:comp_symm}
\end{figure}

The last step in our analysis of the ensemble forecasts of the eight storms is assessing how their kinematic structure and intensity are predicted by examining forecasts of symmetry, compactness and MSLP. Overall, these forecasts also show a non-gradual evolution with lead time, as previously observed for occurrence, position and thermal structure forecasts. Specifically, the forecast median often does not converge gradually and monotonically towards the analysis value, the forecast spread does not always decrease gradually and monotonically and forecast jumps occur at some lead times for most cases. However, the evolution of these forecasts is more gradual than that of the forecasts previously examined. For this reason, we focus here on the overall performance of these forecasts rather than on their evolution with lead time. Full forecast statistics are only shown for two representative cases, namely Numa for compactness (Figure \ref{fig:comp_symm}a) and Stephanie for symmetry (Figure \ref{fig:comp_symm}b).

It is apparent that both compactness and symmetry are somewhat underpredicted in these two cases, though with a convergence of forecast distributions towards the analysis value at short lead times. These two forecasts are representative of compactness and symmetry forecasts for other cases, in that both the underprediction and the convergence at short lead times are seen in most cases. One could naturally expect compactness to be underpredicted to some extent, given the low resolution of the ECMWF ensemble prediction model. However, the clear convergence of forecasts at short lead times (in most cases to an extent that the analysis value is well within the interval of the forecast distribution and close to its median) and the fact that the distribution tails reach or exceed the analysis value even at long lead times indicate that the model is capable of producing high values of compactness. Moreover, compactness and symmetry forecasts appear to be well correlated with each other, so that high values of either metric are associated to high values of the other. We conclude that the underprediction arises because the occurrence of a very symmetric and compact storm is a highly unlikely event and as such it is by nature near the tail of the forecast distribution (especially at long lead times), as observed by \citet{majumdar_2014}. Later forecasts then tend to converge at short lead times as they benefit from improved initial conditions.

Finally, we note that MSLP forecasts, which are overall the most gradually evolving ones with lead time, do not show any remarkable signal and therefore are not shown here. There appears to be a slight tendency to underpredicting MSLP at long lead times for many storms, which is probably due to the low probability of cyclone occurrence in early forecasts. This hypothesis is supported by forecasts of Qendresa, the most intense of the eight storms (see Section \ref{sec:overview}), which consistently and largely underpredict MSLP, although forecast distribution tails reach the analysis value even at long lead times. Qendresa indeed underwent an extremely rapid development \citep[more than 15 hPa pressure drop in 18 hours, see][]{cioni_2018} which appears as highly unlikely especially in early forecasts, even though the probability of cyclone occurrence is high from 7 days in advance (Figure \ref{fig:results_intro}).

\section{Discussion and conclusions} \label{sec:conclusions}

Medicanes have been gaining increasing attention in the research community in the last two decades. These storms constitute a major threat in the Mediterranean region, due to intense winds and rainfall. Although the pathway leading to the formation of Medicanes is by now well known, they remain elusive characters of the Mediterranean climate in that their frequency of occurrence is low and an objective definition has not yet been found. The predictability of Medicanes is also low due to scarce observations over the sea and the interplay between the numerous factors influencing their entire life cycle at multiple spatial and temporal scales. 

In this paper, the predictability of eight southern European tropical-like cyclones, seven of which Medicanes, is analysed by evaluating ECMWF operational ensemble forecasts against operational analysis data. We apply an object-based approach that allows focusing on specific storm features, while tolerating their shifts in time and space to some extent. Each storm is then treated as an object and its forecasts are evaluated using suitable metrics: MSLP, symmetry, compactness, $-V_T^U$ which give a measure respectively of the cyclone's intensity, symmetry, pressure gradient and upper-level thermal structure and therefore well represent tropical-like traits attained during the mature phase of its life cycle. This object-based approach has shown strengths in extracting the most relevant information from the data and value in condensing it into intuitive metrics: for these reasons, it could easily be applied to other types of forecasts and atmospheric features. The DTW technique in particular looks promising for further application in the atmospheric sciences due to its intuitiveness and flexibility in providing a meaningful space-time matching of time series.

Findings reveal that the evolution of ensemble forecasts with lead time is far from gradual, generally differing from the steady convergence towards the analysis value that may be expected. In particular, rapid increases in the probability of cyclone occurrence (\emph{forecast jumps}) are seen in most cases. This behaviour is compatible with the existence of predictability barriers, similar to the spring predictability barrier observed for ENSO \citep{duan_2013}, which would only be overcome when initial conditions adequately represent the variety of factors playing a role in Medicane development at all scales. Cyclone thermal structure forecasts also exhibit a non-gradual evolution in some cases, with the forecast median drifting away from the analysis value and spread increasing with decreasing lead time. However, late forecasts which have been initialized when the storm has already developed tend to be more accurate than earlier forecasts. This supports previous findings of high sensitivity of Medicane simulations to initial conditions \citep{cioni_2016}.

On the other hand, forecasts of cyclone position exhibit a visible tendency to a consistent spatial distribution of cyclone position uncertainty and bias (i.e. a nonzero median position error) between consecutive forecasts, which may be explained by the fact that some regions of the Mediterranean Sea are more conducive to Medicane development than others \citep{tous_environments,cavicchia_2014_clim}, thus favouring the occurrence of the cyclone in the same region between consecutive forecasts. This implies that early forecasts may already contain valuable prognostic information on the cyclone's position during its mature phase.

Unlike other metrics, compactness and symmetry are consistently underpredicted in most cases, especially at long lead times. A marked improvement of these forecasts is however seen at short lead times. In light of these contrasting behaviours, we exclude the presence of any systematic bias that could be expected due to the relatively low resolution of the ECMWF ensemble model. We instead deem the underprediction to be a result of the intrinsically low probability of the occurrence of a Medicane (that is, a highly axisymmetric and compact storm) in early forecasts, which causes it to be found near the tail of the forecast distribution, as observed by \citet{majumdar_2014}. We interpret in the same way a weak tendency to underpredict MSLP that is seen in some cases at long lead times. Considering all metrics, forecasts indicate that the ECMWF ensemble model can adequately reproduce Medicanes in terms of their tropical-like traits, albeit only at relatively short lead times.

The present work paves the way towards an in-depth investigation of the physical mechanisms underlying the features revealed by our analysis, in particular the non-gradual evolution of forecasts with lead time, forecast jumps and the consistent spatial distribution of cyclone position forecasts. A future study will examine the complex interplay between processes at different spatial and temporal scales leading to the formation of a Medicane and its impact on their predictability and the evolution of ensemble forecasts with lead time.

\section*{acknowledgements}

ECMWF is acknowledged for providing the ensemble forecast and operational analysis datasets. The research leading to these results has been carried out within the subproject C3 ``Multi-scale dynamics and predictability of Atlantic Subtropical Cyclones and Medicanes'' of the Transregional Collaborative Research centre SFB/TRR 165 ``Waves to Weather'' funded by the German Research Foundation (DFG).

\bibliography{bibliography}

\end{document}